\documentstyle[preprint,aps]{revtex}

\draft

\begin{document}

\title{Subband energy in two-band $\delta$-doped semiconductors}

\author{F. Dom\'{\i}nguez-Adame}

\address{Departamento de F\'{\i}sica de Materiales, Universidad
Complutense, 28040 Madrid, Spain}

\maketitle

\begin{abstract}

We study electron dynamics in a two-band $\delta$-doped semiconductor
within the envelope-function approximation.  Using a simple
parametrization of the confining potential arising from the ionized
donors in the $\delta$-doping layer, we are able to find exact solutions
of the Dirac-type equation describing the coupling of host bands.  As an
application we then consider Si $\delta$-doped GaAs.  In particular we
find that the ground subband energy scales as a power law of the Si
concentration per unit area in a wide range of doping levels.  In
addition, the coupling of host bands leads to a depression of the
subband energy due to nonparabolicity effects.

\end{abstract}


\narrowtext

\section{Introduction}

Recently attention has been paid to planar- or $\delta$-doped
semiconducting materials \cite{Zrenner} (see Ref.~\onlinecite{Whall} for
a recent review) due to their potential application in ultra-high speed
electronic and opto-electronic devices.  Such doping profiles are
generated by interrupting the crystal growth of the host material and
evaporating the doping impurity during molecular-beam epitaxy.  Under
appropriate growth conditions, excellent confinement of dopant atoms has
been achieved in GaAs:Si \cite{Ploog}, Si:Sb \cite{Zeindl}, and InP:S
\cite{Cheng}.  As a consequence, the doping profile along the growth
direction $z$ can be represented as $N(z)=N_D\delta(z)$, where $N_D$ is
the two-dimensional donor concentration.  The above profile neglects the
random distribution of donors in the $\delta$-doping layer, which is
valid in the high density limit \cite{Kortus}.

The analysis of the resulting electronic structure relies on the
envelope function approximation \cite{Bastard}.  Assuming parabolic
bands, the envelope functions can be calculated by solving the
Schr\"odinger-type equation for the one-electron potential,
corresponding to decoupled host bands.  This approach works fine in
wide-gap semiconductors, provided that conduction- and valence-band
modulations are small.  However a more realistic band-structure is
required in narrow-gap semiconductors or in those devices whose band
modulation is comparable to the magnitude of the gap, mainly due to
nonparabolicity effects.  In $\delta$-doped semiconductors large band
modulation can be attained since it is possible to reach very high
doping levels, typically larger than $10^{13}\,$cm$^{-2}$, thus leading
to quantum confined states deep in the gap.  This situation is even more
dramatic in sawtooth superlattices, which consist of periodic
alternating $n$- and $p$-type $\delta$-doped sheets, separated by
undoped material \cite{SST}.  It is known that two-band models may
successfully describe those nonparabolicity effects.  The aim of this
paper is to obtain exact solutions of two-band Hamiltonians for single
$\delta$-doped layers, without requiring numerical techniques.  As an
illustration of our treatment, we will consider Si $\delta$-doped GaAs
and compare the results with those obtained using simpler approaches.

\section{Model}

For the calculation of the electron dynamics in $\delta$-doped
semiconductors one requires the one-electron potential due to ionized
donors.  In the high concentration limit, several methods has been used,
like the Thomas-Fermi semiclassical approach \cite{Ioratti,Egues,PRB1}
the local density approximation (LDA) \cite{Degani}, and the Hartree
method \cite{Zrenner2,PRB2}.  In these methods, however, the computation
of the one-electron potential relies on numerical techniques, so an
analytical solution for the envelope functions and subband energies is
no longer possible.  However, Gold {\em al.\/}\cite{Gold} have proposed
a closed form of the one-electron potential which brings accurate
results in a wide range of doping levels.  Hereafter we focus on GaAs,
although we should stress that the treatment is completely general.  In
the high concentration limit, which we are interested in, the solution
of the Poisson equation, assuming that the donor concentration presents
a $\delta$-function profile, gives
\cite{Gold}
\begin{equation}
V(z)=-g\exp\left(-\,{|z|\over\alpha}\right),
\label{potential}
\end{equation}
where $\alpha=38.0\,(N_Da^{*2})^{-1/3}\,$\AA, $g=38.1\,(N_D
a^{*2})^{2/3}\,$meV, with the effective Bohr radius $a^*\sim 100\,$\AA\
in GaAs.  This potential shape holds valid whenever $N_D>5\times
10^{10}\,$cm$^{-2}$, for which the exchange interaction on the subband
population and energy is only very weak \cite{largo}.  Hence we can
confidently neglect this effect hereafter.

In the effective-mass $\bf k\cdot p$ approximation, the electronic wave
function is written as a sum of products of band-edge orbitals with
slowly varying envelope-functions.  Keeping only the two nearby bands,
there are two coupled envelope-functions describing the $s$-like
conduction-band and $p$-like valence-band states of the semiconductor,
subject to an effective $2\times 2$ Dirac Hamiltonian.  Assuming that
the band modulation depends only on $z$, the resulting equation for the
envelope-functions in the conduction- and valence-bands can be written
as follows \cite{Callaway,Beresford}
\begin{equation}
\left[ \begin{array}{cc} E_g/2-E+V(z) & -i\hbar v \partial \\
-i\hbar v \partial & -E_g/2-E+V(z) \end{array} \right]
\left( \begin{array}{c} f_c (z)\\ f_v(z)\end{array} \right) =0,
\label{Dirac}
\end{equation}
where $\partial =d/dz$, $E_g=1.42\,$eV is the gap of GaAs and $V(z)$
gives the gap center.  Here the energy is measured from the gap center
at $|z|\to\infty$.  The velocity $v$ is related to the Kane's momentum
matrix elements and is given by $v=\sqrt{E_g/2m^*}$.  In particular,
$\hbar v=9.0\,$eV\AA\ in GaAs.  It should be mentioned that the non-zero
in-plane momentum can be easily absorbed in the parameter definitions
and we will ignore it in what follows.  It is worth mentioning that
electrons and holes are treated within the same footing using this
simple two-band model.  This is so because the effects of other bands
are not included in the Hamiltonian.  However, it is known that
significant spin-orbit coupling takes place in III-V semiconductor.
Thus, a more elaborate treatment should include envelope-functions for
three interactings bands ---conduction, light hole and split-off
hole---.  Fortunately, a detailed treatment of a three-band
semiconductor can be reduced to a two-band semiconductor by a suitable
redefinition of the envelope-functions \cite{Beresford2}.  Hence a
two-band model brings a simple way to study III-V semiconductors.

To solve Eq.~(\ref{Dirac}) we use the {\em ansatz} \cite{Feynman}
\begin{equation}
\left( \begin{array}{c} f_c (z)\\ f_v(z)\end{array} \right)=
\left[ \begin{array}{cc} E_g/2+E-V(z) & -i\hbar v \partial \\
-i\hbar v \partial & -E_g/2+E-V(z) \end{array} \right]
\left( \begin{array}{c} \phi(z)\\ \phi(z)\end{array} \right),
\label{Fey}
\end{equation}
where the function $\phi(z)$ satisfies the equation
\begin{equation}
\left\{ -\hbar^2v^2\,{d^2\phantom{z}\over dz^2} + {E_g^2\over 4}
-[E-V(z)]^2+i\hbar v\,{dV(z)\over dz} \right\}\,\phi(z)=0,
\label{Klein}
\end{equation}
and the potential is given by (\ref{potential}). It is straightforward
although tedious to demonstrate that the solutions can be expressed in
terms of confluent hypergeometric functions $M(a,c;x)$, as defined in
Ref.~\onlinecite{Abra}. However, for the sake of brevity we do not
write down $\phi(z)$ explicitly and simply quote the final result.
Once $\phi(z)$ is known, we can make use of Eq.~(\ref{Fey}) to obtain
the envelope functions. Thus
\begin{mathletters}
\label{final}
\begin{equation}
\left( \begin{array}{c} f_c (z)\\ f_v(z)\end{array} \right)=
A_{+}\exp\left(-\,{zq\over\alpha}+i\xi e^{-z/\alpha}\right)
\left( \begin{array}{c} G_1^*(-z)+{\displaystyle{i\over
2}\,{\epsilon_g\over q-i\epsilon}}\,G_0^*(-z) \\
G_1^*(-z)-{\displaystyle{i\over 2}\,{\epsilon_g\over
q-i\epsilon}}\,G_0^*(-z) \end{array} \right)
\label{finala}
\end{equation}
for $z>0$ and
\begin{equation}
\left( \begin{array}{c} f_c (z)\\ f_v(z)\end{array} \right)=
A_{-}\exp\left({zq\over\alpha}+i\xi e^{z/\alpha}\right)
 \left( \begin{array}{c} G_1(z)-{\displaystyle{i\over
2}\,{\epsilon_g\over q+i\epsilon}}\,G_0(z) \\
G_1(z)+{\displaystyle{i\over 2}\,{\epsilon_g\over
q+i\epsilon}}\,G_0(z) \end{array} \right)
\label{finalb}
\end{equation}
for $z<0$, where $A_{\pm}$ are constants.  For brevity we have defined
the following dimensionless parameters $\epsilon=E\alpha/\hbar v$,
$\epsilon_g=E_g\alpha/\hbar v$, $\xi=g\alpha/\hbar v$ and
$q^2=\epsilon_g^2/4-\epsilon^2$.  The function $G_k(z)$ with $k=0,1$
is defined as follows
\begin{equation}
G_k(z)=M(k+q+i\epsilon,1+2q,2i\xi e^{z/\alpha}).
\label{finalc}
\end{equation}
\end{mathletters}

The energies can be obtained by imposing the continuity of the envelope
functions at $z=0$. In doing so, one finally gets
\begin{equation}
E=\left( {E_g\over 2} \right)\, \cos (\lambda_1-\lambda_0),
\label{energy}
\end{equation}
where $\lambda_k=\lambda_k(E)=\mbox{arg}[G_k(0)]$, with $k=0,1$.  Notice
that $E$ is found by solving a transcendental equation using the usual
search methods.  In the range of parameters we have studied, a very fast
convergence of the confluent hypergeometric series $M(a,b;x)$ is
attained, and thus only the first few terms must be computed.
Therefore, solution of the transcendental equation (\ref{energy}) takes
very short CPU times in most computers.

\section{Results and discussion}

As an illustration, we have calculated the binding energy of the ground
subband $E_b$, defined as the difference in energy between the
conduction-band edge far away from the $\delta$-layer and the ground
electron subband, as a function of the donor concentration $N_D$.  This
binding energy is an important parameter since it can be readily
determined experimentally, for instance, by deep level transient
spectroscopy \cite{Zhu}.  Results are shown in Fig.~\ref{fig1} in a
log-log scale.  As expected, the binding energy increases on increasing
the doping level.  Interestingly, the binding energy is of the form
$E_b=11.184(N_Da^{*2})^{2/3}\,$meV, that is, it scales as $N_D^{2/3}$,
in a similar fashion that $g$ defined above.  Analogous power law is
found in V-shaped potential wells, often considered a good description
of $\delta$-doped semiconductors, by means of ordinary one-band
Hamiltonians within the envelope function approach \cite{Schubert}.
However, Dirac-type Hamiltonians for V-shaped potential well cannot
support quantum confined states because electron states can tunnel
through to hole states, even for small electric fields \cite{Capri,EL}.
This phenomenon is similar to the well-known Klein paradox in quantum
electrodynamics.  Hence the use of (\ref{potential}) overcomes such
difficulties while it retains most of the main features of subband
energy in V-shaped potential (e.g., scaling law with $N_D$).  Finally,
let us comment the results when the same potential (\ref{potential}) is
considered in a one-band framework, i.e., using the standard Ben
Daniel-Duke Hamiltonian \cite{Bastar2}.  After some algebra, the binding
energy of the ground subband $E_b$ can be obtained from the following
transcendental equation involving Bessel functions $J_\nu(x)$
\begin{equation}
J_{\nu+1}(\tilde{g})-J_{\nu-1}(\tilde{g})=0,
\label{Bessel}
\end{equation}
where $\nu=(2\alpha/\hbar v)\sqrt{E_bE_g}\,$ and $\tilde{g}=
(2\alpha/\hbar v)\sqrt{gE_g}\,$.  After finding numerically the roots of
this equation, we obtain that in the one-band model the binding energy
is given by $E_b=11.179(Na^{*2})^{2/3}\,$meV.  Two important points
should be remarked.  Firstly, once again the binding energy scales as
$N_D^{2/3}$.  Secondly, this binding energy is always smaller than that
obtained in the two-band model for the same doping level.  In other
words, the subband in the one-band model is above the subband in the
two-band model.  The depression of levels when coupling of host bands is
considered has also been found in sawtooth GaAs superlattices
\cite{SST}.  The explanation relies on the fact that the nonparabolicity
effects imply an increment of the effective mass with energy and,
consequently, the electronic levels are lowered.  It is clear that the
difference in energy we have found is not very large due to the wide gap
of GaAs ($E_g=1.42\,$eV) as compared to conduction-band modulation,
which amounts up to $0.23\,$eV for the higher doping level considered in
the present study.  Nevertheless, in other materials like Si:Sb, where
the gap is narrower and high doping level can be reached, the difference
should be much larger.

\section{Concluding remarks}

In conclusion, we have presented a theoretical study of $\delta$-doped
semiconductors within the two-band model framework.  As an example, we
have considered Si $\delta$-doped GaAs, for which a number of
theoretical and experimental results are available.  However our present
treatment should be valid in various $\delta$-doped semiconducting
materials.  A simple parametrization of the one-electron potential due
to ionized dopant atoms allows us to find exact solutions for the
envelope function and subband energy.  One of the main conclusions of
the work is that the binding energy scales with $N_D^{2/3}$, similarly
to which is found in solving the Schr\"odinger equation for the ideal
V-shaped potential.  We have also evaluated the effects of the coupling
of host bands and obtained that they lead to a depression of the subband
energy.  This is a direct consequence of nonparabolicity effects, which
manifest themselves via an increase of the effective mass with energy.

\acknowledgments

This work is supported by CICYT (Spain) through project MAT95-0325.

\begin{figure}
\caption{Binding energy in Si $\delta$-doped GaAs as a function of the
donor concentration in a two-band model.}
\label{fig1}
\end{figure}

\end{document}